\documentclass[aps,prb,floatfix,twocolumn]{revtex4}
\usepackage{graphicx}
\usepackage{amsmath}
\usepackage{amssymb}

\newcommand{\BEQ}{\begin{eqnarray}}
\newcommand{\EEQ}{\end{eqnarray}}
\newcommand{\BEA}{\begin{eqnarray}}
\newcommand{\EEA}{\end{eqnarray}}

\renewcommand{\d}{{\rm d}}

\newcommand{\ep}{\varepsilon}

\newcommand{\tr}{{\rm tr}}

\newcommand{\lam}{\varepsilon}

\newcommand{\W}{{\cal W}}

\newcommand{\imaj}{\succ}

\newcommand{\half}{\frac{1}{2}}

\begin{document} 

\title
{Maximal work extraction from quantum systems}
\author{A.E. Allahverdyan$^{1,2)}$, R. Balian$^{3)}$
and Th.M. Nieuwenhuizen$^{1)}$}
\affiliation{$^{1)}$ Institute for Theoretical Physics,
Valckenierstraat 65, 1018 XE Amsterdam, The Netherlands}
\affiliation{$^{2)}$Yerevan Physics Institute,
Alikhanian Brothers St. 2, Yerevan 375036, Armenia}
\affiliation{$^{3)}$ SPhT, CEA-Saclay, 91191 Gif-sur-Yvette cedex, France}

\begin{abstract}

Thermodynamics teaches that if a system initially off-equilibrium is
coupled to work sources, the maximum work that it may yield is
governed by its energy and entropy. For finite systems this bound is
usually not reachable. The maximum extractable work compatible with
quantum mechanics (``ergotropy'') is derived and expressed in terms of
the density matrix and the Hamiltonian.  It is related to the property
of majorization: more major states can provide more work.  Scenarios
of work extraction that contrast the thermodynamic intuition are
discussed, e.g. a state with larger entropy than another may produce
more work, while correlations may increase or reduce the ergotropy.

\end{abstract}



\maketitle

The generality of the laws of thermodynamics for
macroscopic bodies led to a long-lasting effort to derive them from
microphysics \cite{landau,balian,thirring}.  This program is by now
completed. In contrast, for finite systems, the
application and even the formulation of the laws of thermodynamics are
still the subject of studies~\cite{AN,QL2L}.  
The origin of qualitative differences between large and finite systems
was recognized long ago \cite{old1,old2}. 
A quantum system submitted
to time-dependent external potentials that describe work sources
undergoes a unitary transformation. During such an evolution, the
density matrix has constant eigenvalues, thus it cannot
become gibbsian when starting from an arbitrary initial state.  
In contrast, most macroscopic systems have a
thermodynamic behavior: they evolve close to a Gibbs state under
the effect of slowly varying external potentials, keeping nearly constant
entropy \cite{old2,woron}. This holds in general, although their
evolution is Hamiltonian at the microscopic level, owing to the large
and smooth density of states of the spectrum \cite{old2}.
Accordingly, the responses of finite and of infinite systems to
external perturbations differ qualitatively \cite{old1}.  As an
example, processes were studied which would be reversible in
thermodynamics, but entail a specific irreversibility due to the
finite size of the involved system~\cite{sato,satotasaki}.  

The problem we treat here is well known in thermodynamics. It
initiated its birth in 1824: What is the maximal amount of work that
can be extracted from a system S by means of an external source of
work acting cyclically in a thermally isolated process?  
And what are the criteria for comparing
different states with respect to their work-providing ability?  We
find answers to both questions, and show that for finite
systems they qualitatively disagree with the standard thermodynamical
ones.

A specific case in which a finite system S behaves thermodynamically
in regard to its work production is well known. If S is initially
prepared in a gibbsian state, it cannot produce work when coupled to
a source of work through an external variable which returns to its
initial value~\cite{woron,ANThomson,thirring}.  Such a state is thus
called {\it passive}.  We shall therefore be interested in
off-equilibrium initial states. 
A scenario for preparing such states is to take S consisting of
two non-interacting (or weakly interacting) subsystems, 
to set separately the two parts in thermal contact with heat baths at
different temperatures, and then to decouple them from the
baths. Subsequent coupling with a source of work may yield an
amount of work that we wish to evaluate.

{\it The maximal work-extraction problem} is thus posed in the
following way \cite{landau,callen}. Consider a system S which can
exchange work with external macroscopic sources. The evolution of its
density operator $\rho(t)$ is then generated by a Hamiltonian
$H(t)=H+V(t)$, where the time-dependence of $V(t)$ accounts for work
transfer. 
Following Refs.~\cite{landau,callen}, we call {\it
``cyclic''} a process in which S, originally isolated, is
coupled at the time $t=0$ to external sources of work, and decouples
from them at the time $\tau$. Thus, the driving variables of the
sources are cyclic, and the potential $V(t)$ vanishes before $t=0$ and
after $\tau$: $V(0)=V(\tau)=0$. However, S need not return to its
initial state at the time $\tau$.  The system S is thermally isolated
but may involve energy exchanges between its parts. The initial state
$\rho(0)=\rho_0$ and the Hamiltonian $H$ being given, we look for the
maximum work $\W$ that may be extracted from S for arbitrary
$V(t)$. According to the dynamics 
\BEA
\label{1}
{\rm i}\hbar\,\dot{\rho}=[H(t),\rho(t)],
\EEA
the work $\d W={\rm tr}[\,\rho(t)\,\dot{V}(t)\,]\,\d t$
done on S is 
~\cite{landau,balian,callen}
\BEA
\label{3}
W={\rm tr}[\, \rho(\tau)\,H  \,]-E(\rho_0),\quad
E(\rho_0)\equiv {\rm tr}[\, \rho_0\,H \,]\equiv E_{\rm i}.
\EEA
Among all final states $\rho(\tau)$ reached 
from $\rho_0$ under the action of any potential $V(t)$, we are therefore
looking for the one with lowest final energy $E_{\rm f}
={\rm tr}[\,\rho(\tau)\,H\,]$.

The standard answer \cite{landau,callen} relies on the idea that the
final state $\rho(\tau)$ is Gibbsian and that its von Neumann 
entropy $S=-{\rm tr}\,\rho\ln\rho$ 
cannot decrease between the times $0$ and $\tau$.
The maximum value of the work (\ref{3}) is reached when the final
state has the equilibrium form 
\BEA
\label{odu}
\rho(\tau)=\rho_{\rm eq}=\frac{e^{-\beta
H}}{Z}, \qquad Z={\rm tr}\,e^{-\beta H}, 
\EEA
with $\beta>0$ determined by the
equality of the initial and final entropies:
$\ln Z-\beta {\partial\ln Z }/{\partial\beta}=-{\rm tr}\rho_0\ln\rho_0$.
The largest amount of work $-W$ extractable from S is thus
\BEA
\label{oman}
\W_{\rm th}=E(\rho_0)-TS(\rho_0)+T\ln Z,
\EEA
the familiar difference of free energy between 
initial and  final state, both 
evaluated with the final temperature $T$.

{\it Finite systems.}  The above derivation involves two arguments
which call for some discussion.  Following thermodynamical intuition,
we have first stated that the entropy cannot decrease. In fact, the
von Neumann entropy $S(\rho)$ remains constant during the evolution
(\ref{1}). If S is macroscopic, it is the coarse-grained entropy
which can increase. Anyhow, this point
is harmless since the bound (\ref{oman}) corresponds to
constant-entropy processes of the total system.  We also
implicitly assumed that S may be brought into an equilibrium state
$\rho_{\rm eq}$ by means of some evolution (\ref{1}).  For macroscopic
systems this is usually allowed since dissipative processes within S
may occur while it evolves under the influence of the coupling $V(t)$.

For finite systems, the sole action of $V(t)$ is in general
not sufficient to allow reaching at the time $\tau$ a gibbsian state
of the form (\ref{odu}).  Indeed, not only is the entropy $S(\rho)$
conserved during the evolution generated by (\ref{1}), but all the
eigenvalues of $\rho$\cite{old1,old2}. 
In contrast to thermodynamic systems, finite
systems keep memory of their initial state and do not involve any
relaxation mechanism.  One may therefore expect that the maximal amount of 
work $\W$ extracted from S is generally smaller than $\W_{\rm th}$.

More precisely, the evolution (\ref{1}) of $\rho$ is unitary, so that
$\rho(\tau)= U\,\rho_0\, U^\dagger$. We look for the minimum of the
final energy $E_{\rm f}={\rm tr}\, U\,\rho_0\,U^\dagger H$ over all
unitary operators $U$.  We can parameterize the 
variations $\delta U$ of $U$ as $\delta U=XU$, where $X$ is an
arbitrary infinitesimal antihermitian operator. Hence, we find $\delta
E_{\rm f}={\rm tr} (XU\rho_0U^\dagger H-U\rho_0U^\dagger XH)={\rm
tr}X[\rho(\tau),H]$.  The stationarity of $E_{\rm f}$ thus implies
that $\rho(\tau)$ should commute with $H$ and have the same
eigenvalues as $\rho_0$, a condition which replaces (\ref{odu}). In
the spectral resolutions of $\rho_0$ and $H$,
\BEA
\label{gajl}
&&\rho_0=\sum_{j\geq 1} r_j|r_j\rangle\langle r_j|,\qquad
H=\sum_{k\geq 1} \lam_k|\lam_k\rangle\langle \lam_k|,
\EEA
we order the eigenvalues as
\BEA
\label{777}
r_1\geq r_2\geq\cdots,\qquad
\lam_1\leq\lam_2\leq\cdots 
\EEA
The minimum of
$E_{\rm f}$ is then $\sum_{j}r_j\lam_j$, and
it is reached for 
\BEA
\label{kalan}
\rho(\tau)=\sum_jr_j|\lam_j\rangle\langle\lam_j|,
\EEA
which is stationary since it commutes with $H$.  This result is
consistent with the extension for finite systems of the second law in
Thomson's formulation \cite{woron,ANThomson,thirring}: the two
conditions which characterize the state $\rho(\tau)$ (commutation with
$H$ and ordering (\ref{777})) are the ones which ensure that this
state is passive: no further work can be extracted from S after time
$\tau$ by means of cyclic processes.

Note that, if the spectrum of $H$ involves degeneracies which
have no counterpart in $\rho_0$, the final state $\rho(\tau)$ is not
unique (in contrast to $\rho_{\rm eq}$ associated with $\W_{\rm th}$).

Altogether, the maximum of the amount of
work (\ref{3}) that can be extracted from S is
\BEA
\label{ergotropy}
\W=\sum_{j,k}r_j\lam_k\,(\,|\langle r_j|\lam_k\rangle|^2
-\delta_{jk}).
\label{ru}
\EEA
For $\W$, which depends only on the initial state and Hamiltonian, we
coin the name {\it ergotropy} ($'\ep\rho\gamma o\nu$: work;
$\tau\!\rho o\pi\eta'$: transformation, turn).  By construction, we
have $\W_{\rm th}\geq \W\geq 0$. The ergotropy $\W$ vanishes if
$\rho_0$ is passive.  It equals the thermodynamical upper bound
$\W_{\rm th}$ only if there exist two numbers $\beta$ and $Z$ such
that the eigenvalues (\ref{gajl}, \ref{777}) of $\rho_0$ and $H$
satisfy $\ln r_j=-\beta\lam_j-\ln Z$, so as to allow $\rho(\tau)=
U\rho_0U^\dagger$ to reach a gibbsian form (\ref{odu}) in spite of the
lack of thermalization mechanism. Noticeable examples include: 
{\it (i)} a pure initial state of S;
{\it (ii)} two-level systems; 
{\it (iii)} harmonic oscillators in case
$\rho_0$ is a gaussian state, since both the sequences 
$\ep_k$ and $\ln r_j$ are then equidistant.

For macroscopic systems the difference
$\W_{\rm th}-\W$ is 
typically relatively small, since the final
state (\ref{kalan}) may lie close to an equilibrium state if $\tau$ is
large; the spectra of $\ln\rho_0$ and $H$ are dense, and a linear relation
between them is approximately satisfied in the
relevant region.  However, for finite systems 
$\W_{\rm th}- \W$ can be significant. 

It remains to show that the bound $\W$ can actually be reached by
coupling S with some source of work that realizes a cyclic process.
We thus want to find a time $\tau$ and an interaction $V(t)$ which
vanishes at $t=0$ and at $t=\tau$, so that, when added to the Hamiltonian 
$H$, $V(t)$ leads from the initial state $\rho_0$ to $\rho(\tau)$
defined in (\ref{kalan}). 
An evolution operator which realizes this goal is 
\BEA
\label{opto}
U=\sum_j|\ep_j\rangle\langle r_j|. 
\EEA
A Hamiltonian $H+V(t)$ generates in
the interaction representation an evolution operator $U_{\rm I}$
which satisfies 
\BEA
\label{barbarossa}
i\hbar\frac{\d U_{\rm I}(t)}{\d t}=e^{iHt/\hbar}\,V(t)\,e^{-iHt/\hbar}
\,U_{\rm I}(t),
\EEA
with $U_{\rm I}(0)=1$. We define
$U_{\rm I}(\tau)\equiv e^{iH\tau/\hbar}\,U
= e^{-i\Lambda \tau/\hbar}$,
where $\Lambda$ is obtained by diagonalization.
We choose  for
$U_{\rm I}(t)$ the simple form $U_{\rm I}(t)=e^{-i\Lambda
\varphi(t)/\hbar}$, where $\varphi(0)=\dot\varphi(0)=
\dot \varphi(\tau)=0$
and $\varphi(\tau)=\tau$. Then the potential $V(t)=\dot\varphi(t)
e^{-iHt/\hbar}\,\Lambda\,e^{iHt/\hbar}$ 
describes according to (\ref{barbarossa}) 
a source of work that extracts during 
the time $\tau$ the work (\ref{ergotropy}) from S. 
This potential is far from unique.

In case not only $\rho(\tau)$ but also the initial state is
stationary, $[\rho_0,H]=0$, we can choose the same eigenbases for
$\rho_0$ and $H$, but (\ref{777}) implies that $|r_1\rangle,
|r_2\rangle,...$ in (\ref{gajl}) are deduced from $|\lam_1\rangle,
|\lam_2\rangle,...$ by some permutation. Then the matrix $|\langle
r_j|\lam_k\rangle|^2$ in (\ref{ergotropy}) is a permutation matrix.
For instance, if the lowest two levels $\ep_1<\ep_2$ have initially
the inverted populations $r_2<r_1$, respectively, so that
$|r_1\rangle=|\ep_2\rangle$ and $|r_2\rangle=|\ep_1\rangle$, we may
easily implement the  transformation $U$, a permutation 
that interchanges
$r_1$ and $r_2$, either in a rapid or in a quasistatic regime.

{\it Comparison of activities.} We wish to compare two states $\rho_0$
and $\sigma_0$ of a system S as regards the maximum work that they may
provide. To make such a comparison meaningful, we assume the initial
energies to be the same, $E(\rho_0)=E(\sigma_0)$. If S is macroscopic
and can reach equilibrium at the end of the process, $\W_{\rm th}$
depends only on the entropy $S$ of the initial state, and it
decreases when $S$ increases, since $-\d\W_{\rm th}/\d S$ is the
temperature of the final state reached. However, the situation is
different for finite systems. Consider, for instance, a
three-level system with eigenenergies $\lam_{1,3}=\mp 1$, $\lam_2=0$.
The eigenstates of $\rho_0$ and $\sigma_0$ are taken as
$|r_{1,3}\rangle=|s_{1,3}\rangle=
(|\lam_1\rangle\mp|\lam_3\rangle)/\sqrt{2}$,
$|r_{2}\rangle=|s_{2}\rangle=|\lam_2\rangle$, so that $E(\rho_0)=
E(\sigma_0)=0$. If their eigenvalues are $\{r_j\}=\{0.90,0.08,0.02\}$
and $\{s_j\}=\{0.91,0.05,0.04\}$, the entropy $S(\rho_0)\simeq 0.375$
exceeds $S(\sigma_0)\simeq 0.364$. Accordingly, the thermodynamic
bound $\W_{\rm th}(\rho_0)\simeq 0.882$ for the work is smaller than
$\W_{\rm th}(\sigma_0)\simeq 0.887$. Nevertheless, the ergotropy
$\W(\rho_0)=0.88$ of $\rho_0$ is larger than the ergotropy 
$\W(\sigma_0)=0.87$ of $\sigma_0$. 
The actually reachable bounds are, as expected, lower
than the corresponding $\W_{\rm th}$'s, but they are reversed in
order: the entropically more disordered state $\rho_0$ may provide
more work. 

Thus, the entropy criterion fails for comparing the ergotropies.
The theory of majorization \cite{major}, that we briefly
recall now, provides another criterion which may be helpful. 
In quantum statistical mechanics \cite{thirring,fabio} 
a density operator $\rho$ is said to majorize
$\sigma$ if their eigenvalues $r_j$ and $s_j$, 
set in the decreasing order (\ref{777}), satisfy 
\BEA
\label{durs}
\sum_{j=1}^kr_j\geq \sum_{j=1}^ks_j,\qquad
{\rm for~~any}\quad k\geq 1.
\EEA
This property, denoted as $\rho\imaj \sigma$, is transitive 
($\rho\imaj \sigma$ and $\sigma\imaj \tau$ imply $\rho\imaj \tau$).
It characterizes order, but in a stronger way than entropy since
$\rho\imaj \sigma$ implies not only $S(\rho)\leq S(\sigma)$, but
also ${\tr}\,f(\rho)\leq {\tr}\,f(\sigma)$ for any concave function
$f(x)$. 
Pure states majorize all states, while in a Hilbert space of
dimension $n$ all states majorize $\rho=1/n$. If we have both
$\rho\imaj \sigma$ and  $\sigma\imaj\rho$, then $\rho$ and $\sigma$ are
unitarily equivalent. However, the order defined by 
majorization is incomplete
since, for $n\geq 3$, pairs of states $\rho$ and $\sigma$ exist 
of which neither majorizes the other.  

Returning to the comparison of the activities of $\rho_0$ and $\sigma_0$
with $E(\rho_0)=E(\sigma_0)$,
we find after summation by parts the ergotropy difference
$\delta \W\equiv\W(\rho_0)-\W(\sigma_0)$ as
\BEA
\delta\W
=\sum_{j\geq 1}(s_j-r_j)\lam_j
=\sum_{k\geq 1}(\lam_{k+1}-\lam_k)\sum_{j=1}^k(r_j-s_j).
\EEA
Hence a sufficient condition for $\rho_0$ to be more active than
$\sigma_0$ is $\rho_0\imaj\sigma_0$ ($S(\rho_0)\leq S(\sigma_0)$
alone is neither necessary nor sufficient). 
There exists a wide class of non-unitary
evolutions~\cite{fabio} which lead from a state $\rho_0$
to $\sigma_0$ such that $\rho_0\succ\sigma_0$. If $\rho_0$
and $\sigma_0$ have the same energy, we have $\W(\rho_0)\leq
\W(\sigma_0)$.
For instance, the diagonal part
$\bar\rho_0 =\sum_k|\lam_k\rangle\,\langle\lam_k|\rho_0|\lam_k\rangle\,
\langle\lam_k|$ of $\rho_0$ has the same energy as $\rho_0$ itself, but
$\W(\rho_0)\geq \W(\bar\rho_0)$ since $\rho_0\imaj\bar\rho_0$~\cite{major}. 
In order to find opposite behaviors in the comparison of ergotropies
and of free energies (or entropies), we have to search for cases
when $\rho_0\not\prec\sigma_0$ and $\rho_0\not\succ\sigma_0$.
All pairs of inequalities
$\W(\rho_0)\gtrless\W(\sigma_0)$ and $S(\rho_0)\gtrless S(\sigma_0)$
may then occur, as illustrated by examples given above and below.

{\it Auxiliary system.}
If S is supplemented with an auxiliary system ${\rm \Omega}$ with
Hamiltonian $H_{\rm \Omega}$ and initial state $\omega_0$, the overall
Hamiltonian $H+H_{\rm \Omega}+V(t)$, where
$V(t)$ couples ${\rm S+\Omega}$ with external sources,
generates a unitary transformation in the product Hilbert space, which
is more general than for work sources coupled separately to S and
${\rm \Omega}$.  The initial state $\rho_0\otimes\omega_0$ is
uncorrelated and the evolution conserves its (factorized)
eigenvalues. The ergotropy satisfies
$\W(\rho_0\otimes\omega_0)\geq\W(\rho_0)+\W(\omega_0)$, the same
inequality as for $\W_{\rm th}$. 
Consider, however, again two states $\rho_0$ and $\sigma_0$ with the
same energy. If they are macroscopic and satisfy $\W_{\rm
th}(\rho_0)>\W_{\rm th}(\sigma_0)$, this ordering of activities is not
changed by the introduction of the auxiliary system, since the
additivity of entropy for the initial states implies $\W_{\rm
th}(\rho_0\otimes\omega_0)>\W_{\rm th}(\sigma_0\otimes\omega_0)$.
But for finite systems the order of ergotropies can be reversed.
As a first example consider for S the same three-level system as
above, with eigenvalues of $\rho_0$ and $\sigma_0$ now equal to
$\{r_j\}=\{0.8,0.1,0.1\}$ and $\{s_j\}=\{0.5,0.5,0\}$. For ${\rm
\Omega}$ we take a two-level system with 
eigenenergies $0$ and $\Delta>0$, initially in a pure state. 
Although $\rho_0\not\imaj\sigma_0$, we have both $S(\rho_0)\simeq 0.639
<S(\sigma_0)\simeq 0.693$ and $\W(\rho_0)=0.7>\W(\sigma_0)=0.5$.
However, coupling with ${\rm \Omega}$, which does not change the
entropies, reverses the inequality for the ergotropies if
$\Delta<1/4$, since (for $\Delta <1$) 
$\W(\rho_0\otimes\omega_0)
-\W(\sigma_0\otimes\omega_0) =0.4\,\Delta-0.1$. 
For $\Delta=0$, the
auxiliary system ${\rm \Omega}$ does not contribute to the energy
balance but nevertheless raises $\W$ owing to its order.

The opposite situation is also possible, provided S has at least four
levels. Consider for instance a system S with eigenenergies
$\ep_{1,4}=\mp 1$, $\ep_{2,3}=\mp (1-x)$, with $0<x<1$. As eigenstates
of $\rho_0$ and $\sigma_0$ we take
$|r_{1,4}\rangle=|s_{1,4}\rangle=(|\ep_{1}\rangle\pm|\ep_{4}\rangle)
/\sqrt{2}$,
$|r_{2,3}\rangle=|s_{2,3}\rangle=(|\ep_{2}\rangle\pm|\ep_{3}\rangle)
/\sqrt{2}$, which ensures $E(\rho_0)=E(\sigma_0)=0$, and as
eigenvalues:
\BEA
\label{moroz1}
\{r_j\}=\frac{1}{(1+w)^2}\left\{
w(w+3),\,\frac{1-w}{2},\,\frac{1-w}{2},
\, 0\right\},\\
\label{moroz2}
\{s_j\}=\frac{1}{(1+w)^2}\left\{2w,\,2w,\,w(1-w)^2, \,
(1-w)^3\right\}, 
\EEA 
which are ordered according to (\ref{777})
provided $1>w>1/2$.  The fact that $\rho_0\not\imaj\sigma_0$ and
$\sigma_0\not\imaj\rho_0$ allows to violate the thermodynamical
ordering, since we have simultaneously $S(\rho_0)<S(\sigma_0)$ and
$\W(\rho_0)<\W(\sigma_0)$ for sufficiently small $x$, as seen from
$(1+w)^2\,[\W(\sigma_0)-\W(\rho_0)]=(1-w)(2w-1)-xw(1+2w-w^2)$.  Take
${\rm \Omega}$ as a two-level system; its only
relevant feature will be the eigenvalues $\{w,1-w\}$ of its initial
state $\omega_0$. Provided $1/4<w^2< 1/2$, the equations (\ref{durs})
for $k=1,2,...,8$ are satisfied and hence $\rho_0\otimes \omega_0\imaj
\sigma_0\otimes\omega_0$. This implies $\W(\rho_0\otimes \omega_0)>
\W(\sigma_0\otimes\omega_0)$: ${\rm \Omega}$ restores for the
ergotropies the order inferred from the
thermodynamic relation $S(\rho_0)<S(\sigma_0)$.

All these contradictions between the predictions of thermodynamics
and the behavior of finite systems do not occur for the subset of
states that may ordered in the sense of majorization: then
$\rho_0\succ \sigma_0$ implies both $S(\rho_0)\leq S(\sigma_0)$ and
$\W(\rho_0)\geq \W(\sigma_0)$ and $\W(\rho_0\otimes\omega_0)\geq
\W(\sigma_0\otimes\omega_0)$ for an arbitrary $\omega_0$.

{\it Correlations.} In the above examples the initial state of S$+\Omega$
was uncorrelated, but not the final state, because evolution permutes its 
eigenvectors although its eigenvalues remain factorized.  
Thus, contrary to thermodynamic intuition, the maximum work can be 
achieved owing to creation of correlations.

Conversely, if the initial state $Q_0$ of S$+\Omega$ is correlated, 
thermodynamics predicts that the available work is increased,
due to the subadditivity of entropy,
$S(Q_0)\leq S(\rho_0)+S(\sigma_0)$, where $\rho_0={\rm tr}_{\rm
\Omega}Q_0$ and $\sigma_0={\rm tr}_{\rm S}Q_0$ are the marginal states
of S and ${\rm \Omega}$, respectively. However,
for finite quantum systems, we have to compare the ergotropies $\W(Q_0)$ and
$\W(\rho_0\otimes\sigma_0)$. 
Take for S a system with three energy levels $\ep_i$, $i=1,2,3$ and
for ${\rm \Omega}$ a two-level system $k=1,2$ with energy levels
$0,\ep$ such that $0<\ep<\ep_2-\ep_1$, $\ep<\ep_3-\ep_2$, and for
${\rm S+\Omega}$ a stationary initial state $Q_0$. Denoting the common
eigenstates of $H+H_{\rm \Omega}$ and $Q_0$ as $|i,k\rangle$, we
assume that the eigenvalues $q_{ik}$ of $Q_0$ are ordered as
$q_{11}>q_{12}>q_{21}>q_{22}>q_{31}>q_{32}$. Then $Q_0$ is both
correlated and passive. Suppressing correlations leads to a factorized
state $\rho_0\otimes\omega_0$ with the same energy, and eigenvalues
$r_i=\sum_{k=1}^2 q_{ik}$ for $\rho_0$ and $p_k=\sum_{i=1}^3 q_{ik}$
for $\omega_0$. The ordering of the set $r_ip_k$ may now differ from
that of $q_{ik}$; for instance, if $q_{11}$ is close to one and all
other $q_{ik}$'s are small with the same order of magnitude, we have
$r_2p_2<r_3p_1$, and $\W(\rho_0 \otimes\omega_0)=
(\ep_3-\ep_2-\ep)(r_3p_1-r_2p_2)>0$. Suppressing the correlations has
thus let the system ${\rm S+\Omega}$ become active.  Altogether, the
order carried by correlations, although it manifests itself directly
in the entropy, has no relation with increase or decrease of
ergotropy.

{\it Fluctuations of work.} The work $-W$ extracted from S
by means of external sources acting in a cycle was defined by
(\ref{3}) as the expectation value of the difference between the
initial and final energies of S.  
This expectation value can also be observed as an average
\cite{satotasaki} by coupling the macroscopic sources of work to $N\gg
1$ identical systems S operating in parallel, in such a way that their
total Hamiltonian is the sum of the Hamiltonians $H+V(t)$ of each
one. However, for a single system S, the work is expected to fluctuate.
In order to evaluate such fluctuations, we need to define work as a
random variable for the dynamics described by the optimal unitary
transformation (\ref{opto}). To do this, we need to determine, for each
single realization in the statistical ensemble described by
$\rho_0$, the initial and final energy of S. This can be done
unambiguously, without perturbing the overall ensemble $\rho_0$, only
if both $\rho_0$ and $\rho(\tau)$ commute
with $H$. Then the measurement of energy at the time $t=0$ allows to
filter systems which have the same initial energy $\ep^{(\rm i)}_k$
after the measurement.  The evolution operator (\ref{opto}) then
determines for each initial $\ep^{(\rm i)}_k$ the final energy
$\ep^{(\rm f)}_l$, hence the work $\ep^{(\rm i)}_k-\ep^{(\rm f)}_l$
is given by S to the sources. We can thus regard
$\ep^{(\rm i)}_k-\ep^{(\rm f)}_l$ as a {\it classical} random variable
describing work, and governed by the diagonal elements $r_j$ of the
density matrix $\rho_0$.  In the same way as 
$\W=\sum_{jk}r_jP_{jk}(\ep_k-\ep_j)$, where 
$P_{jk}=|\langle \ep_j|U|\ep_k\rangle|^2$ is a permutation matrix, 
we get the {\it ergotropy
dispersion}:
\BEA
\label{dalal}
\Delta \W^2=\sum_{jk}r_jP_{jk}(\ep_k-\ep_j)^2-\W^2.
\EEA
It vanishes for
passive states, for which $P_{jk}=\delta_{jk}$ and $\W\equiv
{\rm max}\, W=0$: one cannot 
hope to put no work in average, and
get some work out by making use of  fluctuations.
For active states, $\Delta \W$ may become sizeable. 
In the example of a two level-system
with the initial state: $\rho_0=r_1|\ep_2\rangle\langle\ep_2|
+r_2|\ep_1\rangle\langle\ep_1|$, the ergotropy is
$\W=(r_1-r_2)(\ep_2-\ep_1)$ and its fluctuation from (\ref{dalal}) is
$\Delta \W$$=2\sqrt{r_1r_2}(\ep_2-\ep_1)$. The relative fluctuation
$\Delta \W/\W$
vanishes for $r_1=1$, $r_2=0$, when $\W$ is largest; 
it becomes large when $\W$ is small,
i.e., $r_1\to r_{2}\to \half$. Many systems
S should then operate in parallel to produce on average the amount $\W$
in a predictable fashion.

{\it Conclusion.}  Maximal work extraction is one of the basic
problems of thermodynamics and has applications in various processes
of energy conversion \cite{landau,callen,QL2L}. In macroscopic
physics, the answer (\ref{oman}) is governed by the non-decrease of
entropy. We have shown that finite devices are less efficient in this
respect: any evolution of a thermally (but not mechanically) isolated
quantum system must leave unchanged not only this entropy, but all the
eigenvalues of the density operator, which prevents in most situations
the thermodynamic bound from being attainable. We have given a general
explicit expression (\ref{ergotropy}) for the ``ergotropy'', the
quantity which, for finite systems, replaces the free energy: it is
the upper bound of the work that a finite system S in a
non-equilibrium initial state $\rho_0$ may yield if it is coupled to
external sources of work undergoing a cyclic transformation.  Many
interaction Hamiltonians $V(t)$ allow to reach this bound.

The proper measure of order for comparing the abilities of work
production of finite systems is thus ergotropy and not free energy.
Several consequences of this result contradict thermodynamic
intuition.  Consider, for instance, a state $\sigma_0$ of S having the
same energy as $\rho_0$ and lower entropy. Thermodynamics suggests
that more work might be extracted from $\sigma_0$ than from $\rho_0$,
and moreover that the presence of an auxiliary system ${\rm \Omega}$,
in a state $\omega_0$ initially uncorrelated with S, preserves this
property.  Such statements can be violated in finite quantum systems.
However, even for finite systems, there is a domain of states defined
by the majorization relation, where predictions of thermodynamics are
qualitatively correct.

This work is supported by the Dutch Science Foundation FOM/NWO.
R. B. acknowledges visits at the UvA.

\end{document}